\documentclass[graybox]{svmult}

\usepackage{mathptmx}       % selects Times Roman as basic font
\usepackage{helvet}         % selects Helvetica as sans-serif font
\usepackage{courier}        % selects Courier as typewriter font
\usepackage{type1cm}        % activate if the above 3 fonts are
\usepackage{makeidx}         % allows index generation
\usepackage{graphicx}        % standard LaTeX graphics tool
\usepackage{multicol}        % used for the two-column index
\usepackage[bottom]{footmisc}% places footnotes at page bottom

\usepackage{amssymb}
\usepackage{todonotes}

\makeindex             % used for the subject index
\begin{document}

\title*{MEDAL: An AI-driven Data Fabric Concept for Elastic Cloud-to-Edge Intelligence}
% Use \titlerunning{Short Title} for an abbreviated version of
% your contribution title if the original one is too long
\author{Vasileios Theodorou, Ilias Gerostathopoulos, Iyad Alshabani, Alberto Abell\'o and David Breitgand}
% Use \authorrunning{Short Title} for an abbreviated version of
% your contribution title if the original one is too long
\institute{Vasileios Theodorou \at Intracom Telecom, Peania, Greece, \email{theovas@intracom-telecom.com}
\and Ilias Gerostathopoulos \at Vrije Universiteit Amsterdam, Amsterdam, Netherlands \email{i.g.gerostathopoulos@vu.nl}
\and Iyad Alshabani \at BitSparkles, Sophia Antipolis, France \email{iyad.alshabani@bitsparkles.com}
\and Alberto Abell\'o \at Universitat Politècnica de Catalunya, Barcelona, Spain \email{aabello@essi.upc.edu}
\and David Breitgand \at IBM, Haifa, Israel \email{davidbr@il.ibm.com}
}
%
% Use the package "url.sty" to avoid
% problems with special characters
% used in your e-mail or web address
%
\maketitle

\abstract*{TBD: Each chapter should be preceded by an abstract (10--15 lines long) that summarizes the content.}

\vspace*{-7em}

\abstract{Current Cloud solutions for Edge Computing are inefficient for data-centric applications, as they focus on the IaaS/PaaS level and they miss the data modeling and operations perspective. Consequently, Edge Computing opportunities are lost due to cumbersome and data assets-agnostic processes for end-to-end deployment over the Cloud-to-Edge continuum. In this paper, we introduce MEDAL---an intelligent Cloud-to-Edge Data Fabric to support Data Operations (DataOps) across the continuum and to automate management and orchestration operations over a combined view of the data and the resource layer. MEDAL facilitates building and managing data workflows on top of existing flexible and composable data services, seamlessly exploiting and federating IaaS/PaaS/SaaS resources across different Cloud and Edge environments. We describe the MEDAL Platform as a usable tool for Data Scientists and Engineers, encompassing our concept and we illustrate its application though a connected cars use case}
\section{Introduction and Motivation}
\label{sec:intro}

Modern consumers seek for personalized innovative services and superior user-experience  that  can  only  be  achieved  through  novel data-driven technologies. Connected Cars, Smart City and Industry 4.0 are notable examples of domains that are backed by mission-critical applications, fueled by and heavily dependent on data. Such applications typically need to process vast amounts of data at various levels to extract actionable information in a timely, reliable and privacy-preserving manner. 
The emergence of Cloud computing has been a huge leap forward to effectively host applications, with on-demand resources and pay-as-you-go business model significantly simplifying management and reducing upfront-investment costs.%, while enabling application ubiquity and the illusion of infinite resources. 

%In the Cloud realm, data-intensive services for Machine Learning (ML) and data analytics are commonly available in a data center acting as monolithic single storage and processing space for various internal and external sources of data. Consequently, all the data needs to be moved to the Cloud before any analysis starts, though resulting in bandwidth costs and unnecessary delays. This model hinders the efficient and timely management of end-user resources and data as well as the latter's quality in the development and maintenance of data applications. 
%\alberto{I do not understand the last part of the sentence, and also it is not clear which is the concrete limitation. I think this is the key issue. Today everybody is working in the Cloud ``without problems". If we want to propose a solution, we first need to explicit which are the concrete problems and show their relevance (if possible, economically quantify the impact). I think introducing "real-time" (or "right-time") is necessary, but not sufficient.}

With the advances in virtualization and cloud-native technologies and the abundance of devices, efficient ways have emerge to store and process data away from centralized data centers and “on-the-Edge”, i.e., closer or even right at the data sources. %According to Gartner [to cite] ``by 2023, more than 50\% of enterprise-generated data will be created and processed outside the data center or Cloud, up from less than 10\% in 2019''. 
This emerging service delivery paradigm referred to as \emph{Edge Computing}, promises to lead to decreased latency, which is of paramount importance for time-critical applications, e.g., autonomous driving, as well as to more efficient utilization of both the current ubiquitous computation resources (smartphones, telecom servers, cars’ on-board units, IoT devices, etc.) and communication bandwidth. Equally importantly, it enables the efficient analysis of data that due to practical, legal, or confidentiality constraints are not allowed to leave the environments in which they were generated, or their transfer entails great performance or other costs.
Edge environments are well represented in Cloud offerings of major Cloud providers, usually as enablers for Internet of Things (IoT) scenarios (AWS IoT Greengrass, Azure IoT Edge, Google Cloud IoT Core).

The problem is that existing Edge computing resources are underutilized, while the network is overutilized. The reason is that although modern Edge offerings support some data preparation and pre-processing services at the Edge, data still needs to be transferred to a central location to be properly analysed. 
Deploying and managing data analytics applications at the Edge is still not straightforward, since existing Cloud solutions for the Edge miss the data modeling and operations perspective. Instead, they focus on an infrastructure and platform level, being oblivious of applications running on top of them.
This view encumbers intelligence from Cloud to the Edge, e.g. decision-making processes of when to move data analytics tasks between Cloud and Edge versus when to move data. 
Thus, existing Cloud solutions for the Edge cannot actively support organizations in the continuous development, operations, and lifecycle management of data analytics applications (DataOps) \cite{ereth-dataops}, which is essential for effectively leveraging data for competitive advantage.
% The process of managing intelligence from Cloud to Edge becomes even more complicated when considering vendor lock-in effects of currently fragmented Cloud offerings.
\textbf{Overall, there is no solution yet that supports advanced DataOps on the Cloud-to-Edge continuum, despite the abundance of mature, yet disconnected, Cloud solutions for data analytics at the Edge.}

To illustrate this problem, we consider the scenario of continuously collecting data from a large number of vehicles and combining them with other context data such as that from wearable sensors and smartphones to detect driving behaviors. The data must be analysed so that statistics over large datasets can be calculated and AI/ML prediction models can be trained to identify correlations and mine frequent patterns. This scenario includes performing anomaly detection to identify different safety-related events, e.g. sudden loss of driver’s focus and scoring of driver behavior. Nevertheless, driver's sensitive data produced within the car may not be allowed to leave the vehicle, or may entail privacy restrictions on being shared among different service providers. In addition, as the number of cars increases, there is a significant rise in the volume of data that needs to be analysed, as well as in the complexity of required data and model management. The challenge then is to deploy, test, execute, and manage service components in the most efficient way, both regarding response time and resource utilization (network bandwidth, compute, storage), from the vehicle to the Cloud, while at the same time respecting data privacy restrictions.

Another challenge is that there is also a methodological gap on how data scientists can deal with the challenges of the Cloud-to-Edge continuum, i.e. the volatility and dynamicity of resources, the varying quality and utility of diverse data sources all along the data path, and the difficulty in discovering and managing relevant data assets \cite{edge-challenges}. 
% Thus, developing a DataOps framework for the Cloud-to-Edge continuum comes with a number of technical challenges:
% To manage dynamicity and quality, service providers need to employ flexible software development approaches that encompass the complete lifecycle of data analytics and ML pipelines. 
% \alberto{There is a huge gap here from Cloud providers to Sw development. What is the relationship between them? Why do you need to improve Sw development to solve the problems above or how this impact the former?}
%The main problem in developing data workflows for the Edge is the extreme distribution, volatility (and mobility) of resources, and the locality, heterogeneity, dynamicity, and varying quality of data. %Current Cloud solutions entail business processes of multiple, fragmented steps, time- and cost-overhead of moving data to centralized data lakes, and risk of data workflows becoming obsolete due to resource volatility and changing requirements (e.g., in data quality). In this challenging setting, end-to-end data applications still need to be built over a unified view over the data and to be able to handle the lifecycle and quality of data analytics. 
A crucial question is how to abstract and obtain a data-centric view of the underlying infrastructure and assets, while at the same time considering the capabilities and opportunities they offer and avoiding vendor lock-in effects. Essentially, the data scientist should be concerned with the data aspects of the analytics workflows, which in turn poses a requirement for sophisticated automation mechanisms to handle and optimize infrastructure and deployment aspects, as well as datasets and data models management and operation, even across operational domains.

\begin{figure}[t]
  \centering
\includegraphics[width=0.4\linewidth]{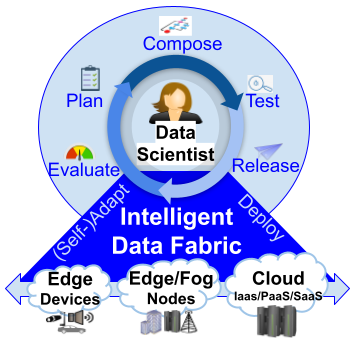}
  \caption{Continuous data application life-cycle management on the cloud-to-edge continuum.}
  \label{fig-vision}
    %\vspace{-5mm}
\end{figure}

\textbf{Our proposed solution aims to support data scientists in the DataOps activities of building and maintaining data analytics applications of high flexibility and quality on the Cloud-to-Edge continuum, optimally utilizing Cloud/Edge resources and services} (Fig.~\ref{fig-vision}).
In particular, it aims to contribute to the evolution of Cloud services for data analytics in the Cloud-to-Edge continuum by: 
% \alberto{The figure should be explained somewhere, expliciting how it addresses each of the challenges above.}

\begin{itemize}
    \item Introducing the MEDAL concept---an \emph{Intelligent Data Fabric} as a continuum on the data application layer, formed by the federation of semantically enabled, cloud-native data-centric constructs acting as building blocks
    \item Offering a platform for AI-driven Cloud-to-Edge DataOps that provides the data scientist with a comprehensive data-centric view over Edge/Fog/Cloud assets, as well as the ability to manage and automate the lifecycle and operation of data-intensive analytics and ML workflows, deployed in a distributed fashion that respects data locality and cost models of data operations.
\end{itemize}

To this end, we introduce innovations in the areas of: (i) Cloud-native Data Fabric across the continuum; (ii) DataOps over the Cloud-to-Edge continuum; (iii) AI-Ops for runtime adaptations over the continuum; and (iv) semantic representation and management of Cloud-to-Edge resources and data assets to ultimately provide a flexible, scalable, and cost-effective platform for Cloud-to-Edge intelligence.

In Sec. 2, we describe the main concepts and methodologies empowering our approach; in Sec. 3, we showcase the application of MEDAL on an illustrative connceted cars use case; finally, in Sec. 4, we conclude this paper with our remarks.

\section{An intelligent Cloud-to-Edge Data Fabric}
\label{sec:concept}

\begin{figure}[t]
  \centering
\includegraphics[width=1\linewidth]{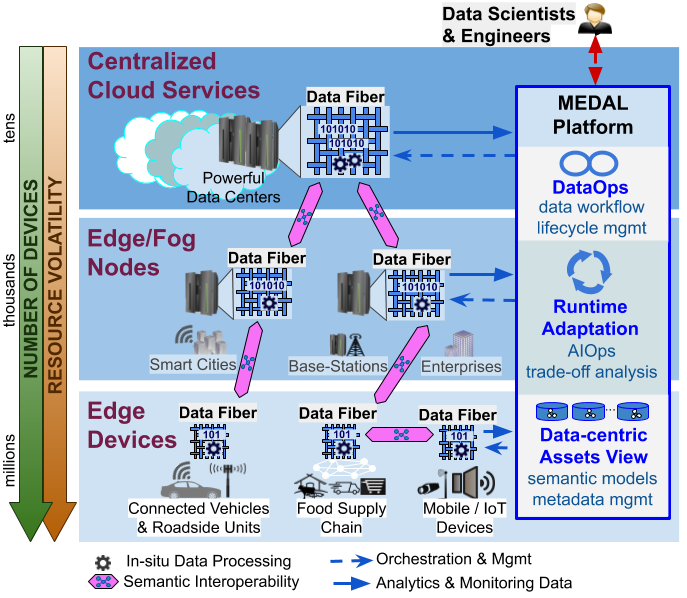}
  \caption{The MEDAL Platform for an intelligent continuum of cloud-native Data Fibers.}
  \label{fig-data-fibers}
    %\vspace{-5mm}
\end{figure}

We adopt a data architectural angle where the main structural component of our data analytics workflow is the \emph{Data Fiber}, which we define as homogeneous wrapper of data assets and services at the data layer. We consider that the \emph{Data Fabric} is formed by the federation of \emph{Data Fibers} of different volumes and capacities on the Cloud-to-Edge continuum (Fig.~\ref{fig-data-fibers}). The Data Fabric facilitates data representation, storage, processing, access and exchange and can be realized using Data Lake technologies \cite{data-lakes} in a distributed manner. The high flexibility and configurability provided by Data Fibers as our structural units primarily stems from following cloud-native principles, according to which data ingestion and state is decoupled from data processing and analytics. This allows for paying the effort and cost of data transformation/integration when it is required, on-demand. 

At a deployment level, Data Fibers across the continuum are realized as containerized micro-services with a focus on scalability and resilience. Data Fibers are equipped with advanced data profiling and summarization mechanisms, as well as with cloud-native capabilities at the resource layer, fostering rapid instantiation of data workflows over collected data where the data resides---a concept also known as in-situ processing \cite{insitu}. Data Fibers may belong to one or more administrative domains (e.g., in multi-cloud setups) and need to interconnect and to interoperate.
Moreover, Data Fibers are highly dynamic and volatile, making it essential to manage their efficient and automated cloud-native orchestration at the infrastructure layer, including primitives such as dynamic provisioning/decommissioning, auto-scaling and migration, trigger-able via declarative interfaces.

We envision the \emph{MEDAL Platform}, a platform to elastically manage and orchestrate Data Fibers and their federations over the continuum, while offering a unified view over underlying data assets and resources to Data Scientists and Engineers (Fig.~\ref{fig-data-fibers}). Thus, the MEDAL Platform composes an intelligent Data Fabric for managing heterogeneous data and resources adaptively, on demand, facing versatile needs and requirements. To achieve these objectives, MEDAL bases on innovative DataOps principles, tools and techniques for managing the complete lifecycle of data applications; AIOps mechanisms for intelligent response to observed events and evolving requirements; and semantic annotation of data assets and metadata management processes, as we further describe in the following subsections.

\subsection{DataOps in the Cloud-to-Edge Continuum}
Recently, the DataOps paradigm has emerged as a catalyst towards data workflow automation, aiming at streamlining data operations, accelerating data application development and fostering quality and continuous improvement throughout all phases of data workflow development and operation \cite{ereth-dataops,capizzi-dataops}. DataOps combines ideas from agile methodologies, DevOps, and lean manufacturing and tries to deal with changing requirements and accelerate time to market, break the silos between development and operations, and improve quality
by reducing non-value-add activities \cite{dataops-book}. DataOps views the development of data analytics as a continuous process and focuses on how to make it iterate faster and with higher quality by advocating both following best practices and using the right tools. 

We tailor the DataOps paradigm and apply it to the development of data analytics in the Cloud-to-Edge continuum. We adopt the ``infinite loop'' of DataOps according to which the development of data analytics passes through different phases: Planning, Composition, Testing, and Release of logical data workflows and Orchestration, Adaptation, and Evaluation of deployed workflows on the continuum. Our DataOps framework includes both (i) methodological principles and best practices that guide data scientists and (ii) tools that help speed up the design and automate the testing, quality assurance, and deployment of data analytics workflows in the continuum. Contrary to other Cloud frameworks and platforms for Edge computing, we support the complete development lifecycle, from design to maintenance, and put emphasis on continuous integration and deployment of data analytics workflows. To this end, the MEDAL Platform, incorporates tools for the following features:

\noindent \textbf{Data Workflow Composition.} MEDAL Platform provides data scientists with customized access to input their data queries as workflows and compositions of different data processing tasks. It exposes (i) visual editors for highly automated development (akin to mashup tools such as Node-RED), and (ii) script/code
editors (akin to Jupyter Notebook) for end-users to directly input their code and define data services. Data Service Composition provides the logical model of a data workflow, which is further mapped by service orchestration tools to a physical model over the available resources. Data workflows can optionally be annotated with requirements of geographical restrictions, resource affinity, capacity (CPU, RAM, throughput), priority and isolation, for optimized mapping.

\noindent \textbf{Monitoring Dashboard.} MEDAL Platform provides visualizations for interactive data and analytics exploration, exposing information about the data analytics outputs and quality at the various application ensembles. In addition, it provides visual graphs for the health, status and availability of infrastructure, as well as log monitoring for event management, as exposed by the Cloud-to-Edge resources.

\noindent \textbf{Autonomic Cloud-to-Edge Management \& Orchestration.} The MEDAL Platform manages flexibility and adaptivity of data workflows as well as provisioning and data asset-aware
coordination of Cloud/Fog/Edge services. In this respect, data workflow deployment (including both service
binding and job scheduling) and quality control are performed in a resource-aware fashion, matching available resources’ characteristics with data workflow requirements. Data
service orchestration can be realised using
workflow and data pipeline management open source tools.% such as Apache Airflow, OpenWhisk and FogFlow.

\noindent \textbf{Continuous Quality Control.} The MEDAL Platform provides data workflow testing and optimization environments for data engineers, for continuous improvement of data and infrastructure compositions. They include mechanisms to create staging environments using Cloud and Edge nodes and test data, and to automate the testing of data workflows in those environments. In particular, input data quality is continuously estimated using the Semantic Knowledge Base described below. Once a data workflow passes its prescribed quality tests, it is deployed in production, where its quality and operation continue to be monitored and profiled.

\subsection{AIOps for Elastic Cloud to Edge Intelligence}

The inclusion of Edge devices, Fog nodes and corresponding services into the pool of Cloud resources introduces new challenges related to volatility, mobility, dynamicity and capacity limitations \cite{edge-challenges}. Advanced IT operations over such complex and dynamic environments are necessary for maintaining quality of deployed data analytics while minimizing resource usage costs. The challenge here is to introduce Edge Intelligence \cite{edge-intelligence} mechanisms both (i) for supporting the elastic lifecycle management and interoperation of Data Fibers (i.e., the data infrastructure layer of our approach) and (ii) for managing the distributed nature of data analytics and ML pipelines spread across the continuum. However, such intelligent mechanisms can only take place with the appropriate visibility and reaction over performance data across all disparate Cloud-to-Edge resources. AIOps \cite{aiops} has recently been proposed as an effective paradigm to exploit AI/ML techniques towards IT operations automation, by correlating data across different interdependent environments and providing real-time, actionable insights over system behaviors, as well as recommendations and (semi-)automated corrective actions. AIOps services provide timely awareness and proactive actions over service quality degradation, resource utilization changes and system mis-configurations, using event management mechanisms combined with application logic to identify root causes and to trigger appropriate restorative management workflows. 

We adopt an AIOps angle of high automation with services of built-in intelligence, where runtime adaptation mechanisms play a central role for closing the loop from issue detection or prediction, to autonomic response. Adaptation mechanisms are crucial for managing unpredictability of resources’ and services’ availability, as well as for accounting for the varying availability and quality of data along the Cloud-to-Edge continuum, which can also continuously change. Runtime adaptation primitives are instilled into proactive management workflows and include: \\
\noindent \textbf{Quality-driven scheduling}: Re-allocation and re-scheduling of data collection and data analytics tasks to sensing/compute nodes based on intelligent monitoring of data and analytics quality;\\
\noindent \textbf{Flexible Data/ML Model deployment}: Move data models across levels (i.e., closer to Cloud or closer to Edge) to efficiently utilize resources and maintain analytics quality, affecting where data aggregation/model training \cite{fl} takes place and thus the necessity of transferring unaggregated/training data across levels;\\
\noindent \textbf{Elasticity of Data Fibers}: Dynamic provisioning, auto-scaling and migration primitives for the Data Fibers across the continuum to respond to detected or predicted over-/under-utilization of resources and to adjust to evolving data analytics requirements (e.g. increase sample size for higher accuracy).

Runtime adaptations follow the Monitor-Analyze-Plan-Execute over Knowledge (MAPE-K) control loop. In the Monitoring phase, data at both the infrastructure and platform level (e.g. CPU load, memory consumption), and at the application level (e.g. application telemetry data on data analytics accuracy and precision, logs). The Analyze phase is responsible for preprocessing, combining, and applying AI/ML techniques for identifying situations that trigger adaptations, also throwing relevant events. Such situations can be both negative (e.g. reduced output quality of a deployed data workflow) and positive (e.g. addition of Edge nodes bringing in opportunity to increase service availability). In the Plan phase, different adaptation actions or plans are determined and compared to each other. If more than one plan is available, a decision is taken either via involving a human operator or (to be fully autonomous) via prioritization based on the contribution of each plan to meeting certain predefined and prioritized goals (e.g. load balancing, increase of output quality). We should note here the importance of cost models \cite{lynceus,adapt-penalties} for the evaluation of alternative plans, which play the role of a Knowledge Base and are continuously augmented with historical data from monitoring and reaction to past events.

Finally, in the Execute phase, the selected plan is rolled out via the activation of a workflow including a series of concrete changes (e.g. provisioning of a new Data Fiber, decommissioning of another one, and starting a computation on the new Data Fiber).

\subsection{Semantic Interoperability in the Cloud-to-Edge Continuum}
Semantic interoperability takes place both at the data layer and the infrastructure resource layer to seamlessly manage heterogeneous resources and services across the Cloud-to-Edge continuum. This can only be achieved with semantically rich information about available computing resources and data assets, combined with appropriate mechanisms for persisting and exchanging such information. In this respect, we introduce the concept of a decentralized \emph{Semantic Knowledge Base} that acts as the source of information used by both (i) management entities to monitor Data Fibers and obtain a unified view over the available resource, data, and service assets; (ii) assets to discover and interoperate with each other. Information includes metadata about infrastructure resource characteristics, data assets (data sources, schemas, profiling, data quality, information available, etc.), monitored runtime state (utilization, active sessions). The Semantic Knowledge Base is also enhanced by predictive cost models for future performance estimations that provide recommendations about the deployment of analytics workflows over the available resources.

To be scalable and allow partially autonomous operation, the Semantic Knowledge Base is decentralized, i.e. there is no central node which keeps track of the metadata in the whole continuum. Instead, nodes form metadata exchange clusters dynamically and only share with other clusters in the continuum the metadata necessary for inter-cluster provisioning and management of data workflows. This way, the single point of failure is avoided and a certain degree of autonomicity in resource management and scheduling of operations is retained by each cluster (which could also be a single node). While the discovery and metadata-exchange process can be continuous, the synapsis of federations between Data Fibers at the continuum can take place on-demand and have a temporal nature.

\section{Application on the Connected Cars Use Case}
\label{sec:usecase}

\begin{figure}[t]
  \centering
\includegraphics[width=\linewidth]{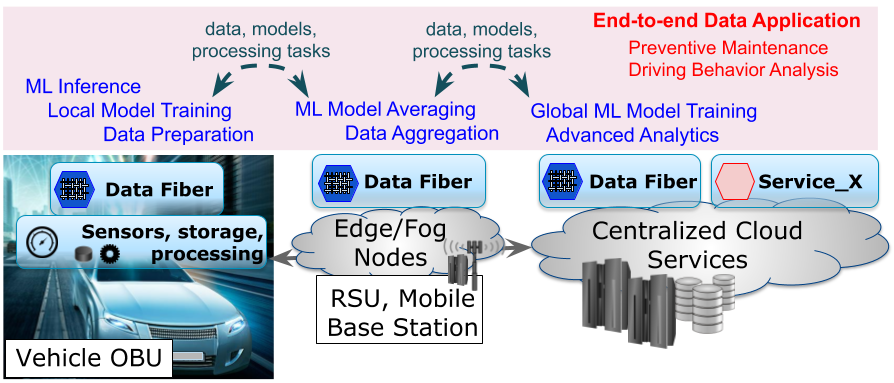}
  \caption{Intelligent Data Fabric for Connected Cars applications.}
  \label{fig-connected-cars}
    %\vspace{-5mm}
\end{figure}

Modern cars are equipped with a plethora of sensors, enabling a variety of services in the context of safety, control and entertainment. Insurance companies, as well as city and road safety administrators and fleet owners, are particularly interested in automatized car analytics such as driving behavior analysis (DBA) and predictive maintenance. %Further examples are to anticipate safety hazards, predict failures, or offer context-aware services to the driver for a rich driving experience.
The value chain ranges from processing units and actuators embedded in the car, to service providers using car data to provide advanced connected services (for the driver, for the manufacturer, for the city, etc.). The execution of analytics over generated data can take place inside the vehicle’s Onboard Units (OBUs), at centralized cloud environments or at intermediary nodes along the Edge to Cloud data path (i.e., Edge/Fog Nodes), such as Roadside Units (RSUs) or cellular network infrastructure (Mobile Base Stations acting as MEC points of presence).

In Fig.~\ref{fig-connected-cars}, we depict the application of the MEDAL concept on this use case. Data Fibers are instantiated through the MEDAL Platform as interconnected containers at the different levels, forming an intelligent Data Fabric. Onboard the car, at the OBU, the Data Fiber collects data from car sensors and performs local storage and processing (i) for \emph{ML inference} tasks such as the diagnosis of hazardous driving behavior or its prediction due to vital signs; (ii) for \emph{local model training} in case of distributed machine learning data applications, e.g., collection of sensitive (DBA) data from thousands of drivers and federated learning of correlations without any raw data actually leaving any car and (iii) for data preparation so that data can be transformed and cleansed accordingly before being moved to higher level Data Fibers. At the Edge/Fog nodes (RSU, Base Station etc.), the Data Fiber performs \emph{model averaging} over the model parameters received from the Data Fibers on the various cars, or \emph{data aggregation}. On the powerful centralized Cloud, the Data Fiber performs global model training and advanced analytics tasks, possibly interoperating with other available services at the Cloud. The scheduling of tasks between different levels, as well as the activation, termination, scaling and migration of Data Fibers, is managed in an automated fashion by the platform, according to availability and cost of resources (e.g., density of cars over a particular geographical area at a particular time) and changing application requirements which are dependent on situational awareness (e.g., spawn Data Fibers in multiple cars close to a traffic collision)

\section{Conclusion}
\label{sec:concl}
In this work, we have introduced MEDAL---a novel concept for the efficient management of the complete lifecycle of data applications deployed all along the Cloud-to-Edge continuum. We constructed the notions for an intelligent Data Fabric composed of Data Fibers---our semantically-enabled cloud-native distributed building units that can dynamically launch, federate and scale on and across the different levels of the Cloud-to-Edge continuum. We described the DataOps, AIOps and semantic annotation principles underpinning MEDAL and we illustrated our approach through a use case from the connected cars domain. In contrast with existing Cloud solutions, MEDAL fully exploits available knowledge about data assets over the continuum and uses this information to provide a unified data and monitoring view to application developers, as well as to make informed decisions about management, orchestration and adaptation of data workflows. As a next step, we plan to build a prototype solution of the MEDAL Platform and conduct large-scale experiments to assess its benefits for interesting distributed learning scenarios.

%\begin{acknowledgement}
%If you want to include acknowledgments of assistance and the like at the end of an individual chapter please use the \verb|acknowledgement| environment -- it will automatically render Springer's preferred layout.
%\end{acknowledgement}
%

%\input{referenc}
\bibliography{medal}
\bibliographystyle{plain}

\end{document}